\documentclass{ws-procs975x65}
\usepackage{amsmath}
\usepackage{amssymb}
\def\vpint{\mathop{\scriptstyle{\mathbf{\diagup}}\hskip-2ex \displaystyle{\int}}}
\font\grb=eurb10
\def\bphi{\hbox{\grb\char'047}\,}

\begin{document}

\title{MONODROMY TRANSFORM APPROACH IN THE THEORY OF INTEGRABLE REDUCTIONS OF EINSTEIN'S FIELD\\ EQUATIONS AND SOME APPLICATIONS\footnote{This research has been partially supported by
the Russian Foundation for Basic Research (grants 05-01-00219, 05-01-00498, 06-01-92057-CE) and the programs "Mathematical Methods of Nonlinear Dynamics" of the Russian Academy of Sciences, and  "Leading Scientific Schools of Russian Federation" (grant NSh-4710.2006.1).}}

\author{GEORGE~ALEKSEEV}

\address{Steklov Mathematical
Institute, Gubkina 8, Moscow 119991, Moscow, Russia\\
\email{G.A.Alekseev@mi.ras.ru}}

\begin{abstract}
A brief sketch of the formulation of the monodromy transform approach and corresponding integral equation methods as well as of various  applications of this approach for solution of integrable symmetry reductions of Einstein's field equations is presented.
\end{abstract}

\bodymatter

\section{Introduction}\label{intro}
For various nonlinear systems integrable by the well known Inverse Scattering Method (called sometimes also the Scattering Transform), the spaces of solutions are parameterized in terms of the scattering data of the corresponding potentials in the associated Schr\"odinger-like equation (associated spectral problem). The scattering data consist of a set of coordinate independent functions of a spectral parameter which characterize uniquely every potential (solution) and which can serve as the "coordinates" in the space of solutions of a given completely integrable system.

In some physically important cases of the symmetry reduced Einstein equations, the spaces of {\it local} solutions also can be parameterized by a finite set of coordinate independent functions of a complex ("spectral") parameter $w$, which determine the branching (monodromy) properties of a fundamental solution of associated linear systems. These data exist for any local solution and thus, in the infinite-dimensional space of local solutions we have two systems of "coordinates" -- the sets of functional parameters whose particular values characterize every local solution uniquely:
\[\left.\begin{array}{l}
\text{the field components:}\\
g_{ik}(x^1,x^2),\,A_i(x^1,x^2),\,\ldots
\end{array}\qquad\right\Vert\qquad
\begin{array}{l}
\text{the monodromy data:}\\
\mathbf{u}_\pm(w),\,\mathbf{v}_\pm(w),\,\ldots
\end{array}
\]
The key difference between these "coordinates" is that the field components should satisfy the field equations, while the space of  monodromy data functions is unconstraint: for arbitrarily chosen set of these functions there exists a uniquely determined local solution of the field equations. The  "coordinate transformation" from the monodromy data to the field components  effectively solves the field equations. That is why we call the approach using this transformation for solution of  symmetry reduced Einstein equations as the "monodromy transform" approach.

The construction of the monodromy transform \cite{Alekseev:1985} provides a unified general base for solving of various integrable symmetry reductions of Einstein's field equations including the Einstein equations for vacuum, the Einstein - Maxwell and the Einstein - Maxwell - Weyl equations for gravitational, electromagnetic and classical neutrino fields as well as for the Einstein equations in higher dimensions which determine the low-energy dynamics of the bosonic sector of some string gravity models \cite{Alekseev:2005b}.

A large variety of physically different types of field configurations can be considered in the framework of this approach. These include the stationary axisymmetric fields of compact sources or asymptotically non-flat fields describing the interaction of these sources with various external fields, the fields of accelerated sources with boost-rotation or boost-translation symmetries, various wave fields such as colliding and nonlinearly interacting waves with smooth profiles or some discontinuities on the wavefronts and having plane, spherical, cylindrical, toroidal or some other forms of the fronts, as well as different inhomogeneous cosmological models with two commuting spatial symmetries. Below we outline some key-points of the monodromy transform approach and mention some its applications.

\section{Parameterization of the solution space by monodromy data}
For electrovacuum Einstein - Maxwell fields depending on two coordinates, any local solution with the complex Ernst potentials $\mathcal{E}$ and $\Phi$, is characterized uniquely by the monodromy data which consist of the four functions of the spectral parameter $w$ holomorphic in some local regions of the spectral plane:
\begin{equation}\label{MData}
\{\mathcal{E}(x^1,x^2),\Phi(x^1,x^2)\}\quad\longleftrightarrow
\quad\{\mathbf{u}_\pm (w),\mathbf{v}_\pm (w)\}
\end{equation}
For vacuum fields $\Phi(x^1,x^2)\equiv 0$ $\leftrightarrow$ $\mathbf{v}_\pm (w)\equiv 0$ and the space of solutions is parameterized by the monodromy data which consist of two arbitrary holomorphic functions $\mathbf{u}_\pm (w)$. For the structure of the monodromy data for other fields see \cite{Alekseev:1985}${}^,$\cite{Alekseev:2005b}{}. To determine the monodromy data for given solution of Einstein equations, one should solve an overdetermined linear system of differential equations whose coefficients depend on the field components of a given solution and their first derivatives.

\enlargethispage*{6pt}
\section{Constructing solutions for arbitrary monodromy data}
\enlargethispage*{6pt}
All components and potentials of a general local solution of electrovacuum Einstein - Maxwell equations can be expressed in quadratures in terms of the monodromy data (\ref{MData}) and of the corresponding solution of a master system of linear singular integral equations whose kernels and rhs are expressed algebraically in terms of
the monodromy data. In particular, given monodromy data, the Ernst potentials are
\begin{equation}\label{Potentials}\mathcal{E}(x^1,x^2)=\epsilon_o-
\displaystyle\int\limits_{L} [\lambda]_{\zeta} k(\zeta)
\bphi^{[\mathbf{u}]}(\zeta)d\zeta,\qquad
\Phi(x^1,x^2)=
\displaystyle\int\limits_{L} [\lambda]_{\zeta}k(\zeta)
\bphi^{[\mathbf{v}]}(\zeta)d\zeta
\end{equation}
where $\epsilon_o=\pm1$ is the value of $\mathcal{E}$ at some initial point; $\zeta\in L$ and the contour $L$ on the spectral plane consists of two disconnected parts $L_+$ and $L_-$ with the endpoints $(\xi_o,\xi)$ and $(\eta_o,\eta)$  depending on the coordinates $x^1$, $x^2$ and coordinates of a chosen initial point; the value $[\lambda]_{\zeta}$ is a jump at the point $\zeta\in L$ of a "standard" branching function $\lambda=\sqrt{(\zeta-\xi)(\zeta-\eta)/(\zeta-\xi_o)(\zeta-\eta_o)}$ and the "weight"  $(\pi/2)k(\zeta)\equiv 1+i\epsilon_0 \zeta {\bf u}^\dagger(\zeta)$, with $\mathbf{u}^\dagger(\zeta)=\overline{\mathbf{u}(\overline{\zeta})}$;  the functions $\bphi^{[\mathbf{u}]}(\zeta)\equiv \bphi^{[\mathbf{u}]}(x^1,x^2,\zeta)$, $\bphi^{[\mathbf{v}]}(\zeta)\equiv \bphi^{[\mathbf{v}]}(x^1,x^2,\zeta)$ should satisfy the linear singular integral equations with the same scalar kernel and different rhs, both depending on the monodromy data ($\tau,\zeta\in L$)\cite{Alekseev:1985}{}:
\begin{equation}\label{LINEs}
\dfrac 1{\pi i}
\vpint\limits_L\dfrac {\mathcal{K}(x^1,x^2,\tau,\zeta)}{\zeta-\tau}\,
\begin{pmatrix}
\bphi^{[\mathbf{u}]}(x^1,x^2,\zeta)\\
\bphi^{[\mathbf{v}]}(x^1,x^2,\zeta)\end{pmatrix}\,
d\zeta=\begin{pmatrix}\mathbf{u}(\tau)\\
\mathbf{v}(\tau)\end{pmatrix}
\end{equation}
Actually, each of these equations in general is a coupled pair of two integral equations, because each function on the disconnected parts $L_\pm$ of the contour is represented by two indpendent functions, e.g. $\mathbf{u}(\tau)$ should be understood as $\mathbf{u}(\tau)=\mathbf{u}_+(\tau)$ for $\tau\in L_+$ and $\mathbf{u}(\tau)=\mathbf{u}_-(\tau)$ for $\tau\in L_-$ and the same is for $\mathbf{v}(\tau)$, $\bphi^{[\mathbf{u}]}(\zeta)$, $\bphi^{[\mathbf{v}]}(\zeta)$.

For different problems the master integral equations (\ref{LINEs}) admit useful modifications. For stationary axisymmetric fields, the regularity axis condition implies $\mathbf{u}_+(\tau)\equiv\mathbf{u}_-(\tau)$, $\mathbf{v}_+(\tau)\equiv\mathbf{v}_-(\tau)$, and therefore, $\bphi^{[\mathbf{u}]}_+(\zeta)=\bphi^{[\mathbf{u}]}_-(\zeta)$,
$\bphi^{[\mathbf{v}]}_+(\zeta)=\bphi^{[\mathbf{v}]}_-(\zeta)$. This allows to merge $L_+$ and $L_-$ and reduce (\ref{LINEs}) to a simple scalar form similar to Sibgatullin's modification of  the Hauser-Ernst integral equations. For the hyperbolic case, (\ref{LINEs}) can be reduced to the quasi-Fredholm integral "evolution equations"  well adapted for solving of the characteristic initial value problems \cite{Alekseev:2001a}.

\section{Applications}

For all of gravitationally interacting fields and for each type of field configurations mentioned in the Introduction, the developed approach suggests the effective tools for analysis of the structure of the whole space of local solutions, a comparison of different solution generating techniques (see, e.g.,\cite{Alekseev:2001c}), construction of infinite hierarchies of exact solutions with arbitrary finite number of free parameters including multi-parametric generalizations and analytical continuations of many known solutions in the space of their parameters \cite{Alekseev:1992}${}^-$\cite{Alekseev-Griffiths:2000a}{}, analysis of asymptotical behaviour of some classes of fields, solution of the Cauchy and characteristic initial value problems for hyperbolic cases \cite{Alekseev:2001a}${}^,$
\cite{Alekseev-Griffiths:2004}{} as well as of the boundary value problems for elliptic cases of integrable reductions of Einstein's field equations. It is clear, however, that in all of the directions outlined above a further work is necessary for the searches of new interesting developments of these methods and their practical applications for solving of various physically interesting problems.
\enlargethispage*{6pt}


\begin{thebibliography}{00}
\bibitem{Alekseev:1985} G.A.Alekseev, {\it Sov.Phys.Dokl.} {\bf 30}, 565 (1985); {\it Proc. Steklov Math. Inst.},  American Math. Soc., {\bf 3}, 215 (1988); {\it Theor. Math. Phys.} {\bf 143}, 720 (2005); gr-qc/0503043.
\bibitem{Alekseev:2005b} G.A. Alekseev, {\it Theor. Math. Phys.} {\bf 144}, 1065 (2005); hep-th/0410246.
\bibitem{Alekseev:2001a} G.A. Alekseev, {\it Theor. Math. Phys.}
{\bf 129}, 1466 (2001); gr-qc/0105111.
\bibitem{Alekseev:2001c} G.A. Alekseev, {\it Physica D} {\bf 152},  97 (2001); gr-qc/0001012.
\bibitem{Alekseev:1992} G.A. Alekseev, Abstracts of GR13 Int. Conf., Cordoba, Argentina,  3 (1992).
\bibitem{Alekseev-Garcia:1996} G.A. Alekseev and A.A. Garcia, Phys.Rev. {\bf D53}, 1853 (1996).
\bibitem{Alekseev-Griffiths:2000a} G.A. Alekseev and J.B. Griffiths, Phys.Rev.Lett. {\bf 84}, 5247 (2000).
\bibitem{Alekseev-Griffiths:2004} G.A. Alekseev and J. B.
Griffiths, Class. Quantum Grav. {\bf 21},  5623 (2004).
\end{thebibliography}
\end{document}